\documentclass[aps,11pt,prd,notitlepage,tightenlines,nofootinbib,showpacs,superscriptaddress]{revtex4-1}
\usepackage{amsmath, amssymb, amsthm}
\usepackage{mathtools}
\usepackage{mathrsfs}
\usepackage{bm}
\usepackage{slashed}   
\usepackage{ulem}
\usepackage{graphicx}
\usepackage{multirow}
\usepackage{tikz}
\usepackage[caption=false]{subfig}
\usepackage{relsize}	
\usepackage{array}
\usepackage{float}
\usepackage{color}
\usepackage{xcolor}
\usepackage{soul}
\usepackage{hyperref}
\hypersetup{colorlinks=true,linkcolor=blue,anchorcolor=blue,citecolor=magenta, filecolor=blue,urlcolor=blue,bookmarksnumbered=true,
pdfview=FitB
}

\def\dd{{\mathrm{d}}}

\mathchardef\-="2D

\newcommand{\half}[1][1] {\mathsmaller{\frac{#1}{2}}}

\colorlet{darkgreen}{green!60!black}
\colorlet{brightyellow}{yellow!75!red}
\colorlet{orange}{red!50!yellow}
\colorlet{darkblue}{blue!60!black}
\colorlet{darkred}{red!80!black}
\colorlet{greenblue}{green!50!blue}

\makeatletter

\newcommand{\Rmnum}[1]{\expandafter\@slowromancap\romannumeral #1@}
\makeatother

\begin{document}
\title{Longitudinal dynamics for mesons on the light cone}

\author{Yang Li}
\affiliation{Department of Modern Physics, University of Science and Technology of China, Hefei, Anhui 230026, China}

\author{James P. Vary}
\affiliation{Department of Physics and Astronomy, Iowa State University, Ames, IA 50011, USA}

\date{\today}
\begin{abstract}
We survey a set of proposed light-front models for confinement in the valence quark sector of the mesons and portray similarities as well as differences.  We present the spectroscopies for the light mesons that result from a selection of longitudinal confinement forms. We note that the Sturm-Liouville theory provides a unifying framework for many elements of this comparison. 
\end{abstract}

\maketitle

\section{Introduction}\label{sect:introduction}

The light-front Schrödinger wave equation (LFSWE) provides a relativistic semiclassical first approximation to QCD for mesons represented as quark and antiquark bound states \cite{deTeramond:2008ht},
\begin{equation}
\Big[\frac{\vec k^2_\perp+m_q^2}{x}+\frac{\vec k^2_\perp+m^2_{\bar q}}{1-x} + V \Big]\psi(x, \vec k_\perp) = M^2 \psi(x, \vec k_\perp). 
\end{equation}
Here, the effective potential $V$ encodes the non-perturbative dynamics of QCD at low energy resolution. 
Remarkably, Brodsky and de Téramond discovered that the LFSWE in the chiral limit ($m_q = m_{\bar q} = 0$) can be identified with the equation of motion of strings in the fifth dimension of Anti-de Sitter (AdS${}_5$) space thus establishing an extraordinary connection between light-front QCD and AdS/QCD, a bottom-up approach known as light-front holography (LFH, \cite{Brodsky:2006uqa}). LFH has been successfully applied to describing the hadron spectroscopy, including the spectra of mesons \cite{Brodsky:2006uqa, Dosch:2015bca, Dosch:2016zdv, Nielsen:2018ytt, Zou:2019tpo},  the infrared behavior of the strong coupling \cite{Deur:2016tte}, hadron form factors \cite{Brodsky:2006uqa, Brodsky:2007hb, Brodsky:2008pf, Sufian:2016hwn} and parton distributions \cite{deTeramond:2018ecg, Liu:2019vsn}. An excellent review of this approach is Ref.~\cite{Brodsky:2014yha}.

 Light-front holography addresses the chiral limit where only the dynamics of the transverse degrees of freedom (d.o.f.) are considered \cite{Li:2022ytx}. Recently, attention has focused on the role of longitudinal dynamics \cite{Chabysheva:2012fe, Glazek:2013jba, Trawinski:2014msa, Li:2015zda, Li:2021jqb, DeTeramond:2021jnn, Ahmady:2021lsh, Ahmady:2021yzh, Shuryak:2021hng, Shuryak:2021mlh, Shuryak:2022thi}. The longitudinal dynamics is important for incorporating finite quark masses, chiral dynamics and longitudinal excitations as well as for identifying the physical states in the excitation spectrum \cite{Gutsche:2012ez}. Unlike the transverse d.o.f., the longitudinal dynamics in the presence of transverse dynamics of the LFSWE has not been posed from underlying principles. 
 
In light of various proposals in the literature, it is tempting to compare these longitudinal confining interactions. Ahmady et al. compared the hadron spectra employing the 't Hooft model and our model \cite{Ahmady:2021lsh}. Shuryak \& Zahed compared the effective light-front Hamiltonian obtained from the Nambu-Goto string having massive ends with a phenomenological model we (with others) introduced \cite{Shuryak:2021hng}. Weller \& Miller further compared the longitudinal confining potentials of these three models and the corresponding wave functions \cite{Weller:2021wog}. 
In this work, we discuss the role of longitudinal dynamics in LFSWE. In particular, we focus on the endpoint behavior of the wave functions and the scaling of the eigenvalues. We show that this behavior is closely related to the singularities of the corresponding LFSWE. This observation
enables us to consider a broader class of longitudinal confining potentials based on the Sturm-Liouville theory. 
We present several concrete examples. 

This perspective is consistent with the traditional quantum many-body approach to self-bound systems, e.g., atoms, molecules and nuclei. There, first approximations are employed to chart out a set of ``orbits'' or ``shells'' of the system. Then, the wave functions of these orbits are adopted as the basis within which the full Hamiltonian operator is diagonalized to obtain the ``exact'' results. In light of this link to the quantum many-body treatment, basis light-front quantization (BLFQ \cite{Vary:2009gt}), we referred to our recent LFSWE approach for light mesons as BLFQ${}_0$.
In the heavy quark systems, the non-relativistic potential model is a valid first approximation \cite{Godfrey:1985xj}. This model is augmented to the Hamiltonian QCD in Coulomb gauge to address the full QCD dynamics \cite{Szczepaniak:1995cw} (see Ref.~\cite{Reinhardt:2017pyr} for a recent review).In the light hadron sector, light-front holography plays an analogous role with its remarkable phenomenological successes. Recently, it was argued that the enpoint asymptotics of the longitudinal dynamics is essential for 
implementing the Gell-Mann-Oakes-Renner (GMOR) relation -- a direct consequence of chiral symmetry breaking in the light sector \cite{Gutsche:2012ez, Li:2021jqb}. 

The reminder of this work is organized as follows. Sect~\ref{sect:formalism} introduces the LFSWE within the separation of variables ansatz and its application in QCD, the light-front holographic QCD. We then discuss the need for a longitudinal dynamics and compare various proposals from the literature in Sect.~\ref{sect:longitudinal_confinement}. In the next section, Sect.~\ref{sect:sturm-liouville_theory}, we construct a general class of longitudinal potentials based on the Sturm-Liouville theory. Several specific cases are investigated. Finally,  we conclude in Sect.~\ref{sect:conclusion}. 

\section{Formalism}\label{sect:formalism} 

Formally, the LFSWE is a low energy effective theory of QCD.  Therefore, it may be obtained from a Hamiltonian renormalization group evolution, either the Okubo-Suzuki-Lee type \cite{Okubo:1954, Suzuki:1980, Suzuki:1982a, Suzuki:1982b, Suzuki:1983, Suzuki:1994} or the flow equation type \cite{Glazek:1993rc, Glazek:1994qc, Wegner:1994}. Alternatively, it can be viewed as the truncation up to the valence Fock sector hence the Tamm-Dancoff type of approximation is applied \cite{Lepage:1980fj, Perry:1990mz}. These formal connections to QCD make it possible to systematically improve the LFSWE, as shown in Refs.~\cite{Wilson:1994fk, Glazek:2012qj, Glazek:2017rwe, Serafin:2018aih}. 
In either approach, one can write down the wave equation, 
 \begin{multline}
\Big[ \frac{\vec k_\perp^2+m^2_q+\Sigma_s(x, \vec k_\perp)}{x} + \frac{\vec k_\perp^2+m^2_{\bar q}+\Sigma_{\bar s}(1-x, -\vec k_\perp)}{1-x} \Big] \psi_{s \bar s/h}^{(J,m_J)}  (x, \vec k_\perp) \\
+ \sum_{s', \bar s'} \int\frac{\dd x'}{2x'(1-x')}\int\frac{\dd^2 k_\perp'}{(2\pi)^3} V_{s\bar ss'\bar s'} (x, \vec k_\perp, x', \vec k'_\perp) \psi_{s'\bar s'/h}^{(J,m_J)} (x', \vec k'_\perp) 
= M^2_h \psi_{s \bar s/h}^{(J,m_J)}  (x, \vec k_\perp).
\end{multline}
$\Sigma_\sigma$ is the quark self-energy and needs to be solved from the gap equation. To the first approximation, we can absorb the self-energy in 
an effective quark mass. $J$ is the total angular momentum and $m_J$ is its magnetic projection. 
$\psi$ is the wave function which is normalized according to,
\begin{equation}
\sum_{s\bar s} \int_0^1\frac{\dd x}{2x(1-x)}\int\frac{\dd^2k_\perp}{(2\pi)^3} \bar\psi_{s\bar s/h}^{(Jm_J)}(x, \vec k_\perp)\psi_{s\bar s/h'}^{(J'm_J')}(x, \vec k_\perp) = \delta_{hh'}.
\end{equation}
Subscript $h$ indicates the hadron species. This wave equation corresponds to the effective Hamiltonian $H_\mathrm{eff} = ({\vec k_\perp^2+m^2_q})/{x} + ({\vec k_\perp^2+m^2_{\bar q}})/({1-x}) + V$. The normalization adopted here stems from the Lorentz invariant phase space
element,
\begin{equation}
\int \frac{\dd^4p}{(2\pi)^4} 2\pi\delta(p^2-m^2) = \int \frac{\dd^3p}{(2\pi)^32p^0}\theta(p^0) = \int \frac{\dd p^+\dd^2p_\perp}{(2\pi)^32p^+} \theta(p^+)
\end{equation}

To associate the eigenvalues and eigenfunctions to hadrons, we need to compute the discrete symmetries $P$ and $C$. They are represented as \cite{Brodsky:2006ez}, 
\begin{align}
C =\;& \int_0^1 \frac{\dd x}{2x(1-x)} \int \frac{\dd^2k_\perp}{(2\pi)^3} \psi^*_{s\bar s}(x, \vec k_\perp) \psi_{\bar s s}(1-x, -\vec k_\perp), \label{eqn:C}\\
(-1)^JP=\;&  \int_0^1 \frac{\dd x}{2x(1-x)} \int \frac{\dd^2k_\perp}{(2\pi)^3} \psi^*_{s\bar s}(x, \vec k_\perp) \psi_{-s-\bar s}(x, \tilde{k}_\perp), \label{eqn:mP}
\end{align}
where $\tilde k_\perp = (-k_x, k_y)$. The mirror parity $(-1)^JP$ is employed here since it is a more convenient observable on the light front \cite{Brodsky:2006ez}. Since the mirror parity flips the sign of the spin, we can use the $m_J=0$ state to compute the mirror parity. 

\subsection{Separation of variables}\label{sect:separation_of_variables}

\begin{figure}
\centering
\subfloat[\  \label{fig:LFQCD_interactions:a}]{\includegraphics[height=0.09\textheight]{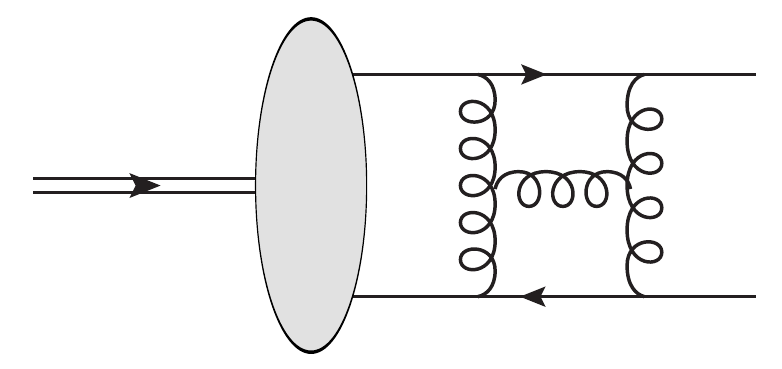}}\qquad
\subfloat[\  \label{fig:LFQCD_interactions:b}]{\includegraphics[height=0.09\textheight]{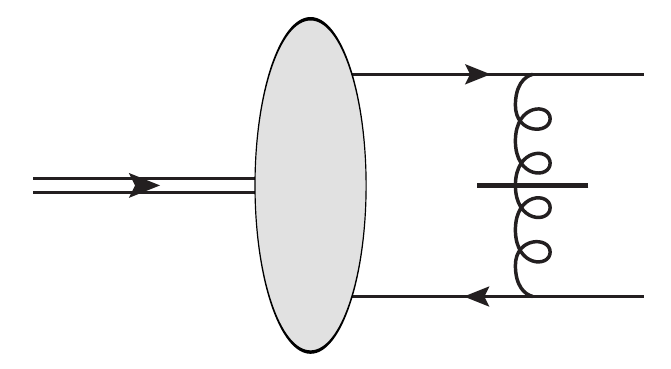}}
\caption{Representative light-front QCD interactions contributed to the (a) transverse and (b) longitudinal effective interactions, respectively.}
\label{fig:LFQCD_interactions}
\end{figure}

In the massless limit, the kinetic energy becomes ${\vec k_\perp^2}/{x(1-x)}$, which depends only on a 2D variable $\vec \kappa_\perp \equiv 
{\vec k_\perp}/{\sqrt{x(1-x)}}$. It is natural to assume the system in the chiral limit also depends only on $\vec \kappa_\perp$ and its conjugate coordinate $\vec \zeta_\perp
= i\nabla_{\kappa_\perp}= \sqrt{x(1-x)}\vec r_\perp$ \cite{Brodsky:2006uqa}. In the more general case, especially when the quark masses are not zero, we can assume that the interaction is separable: $V = V_\perp + V_\|$, an ansatz first explicitly introduced by Chabysheva and Hiller \cite{Chabysheva:2012fe} (cf. Ref.~\cite{Brodsky:2008pf}). We note that the separation of the transverse and longitudinal dynamics is natural in light-front QCD \cite{Brodsky:1997de}. Fig.~\ref{fig:LFQCD_interactions} shows two representative light-front QCD interactions. Fig.~\ref{fig:LFQCD_interactions:a} involves the exchange of the purely transverse gluons while Fig.~\ref{fig:LFQCD_interactions:b} involves the instantaneous interaction in the longitudinal direction. The latter survives in 1+1D as the Schwinger-'t Hooft type of interaction \cite{Schwinger:1962tp, tHooft:1974pnl, Mo:1992sv}.

\begin{figure}
\centering
\includegraphics[width=0.7\textwidth]{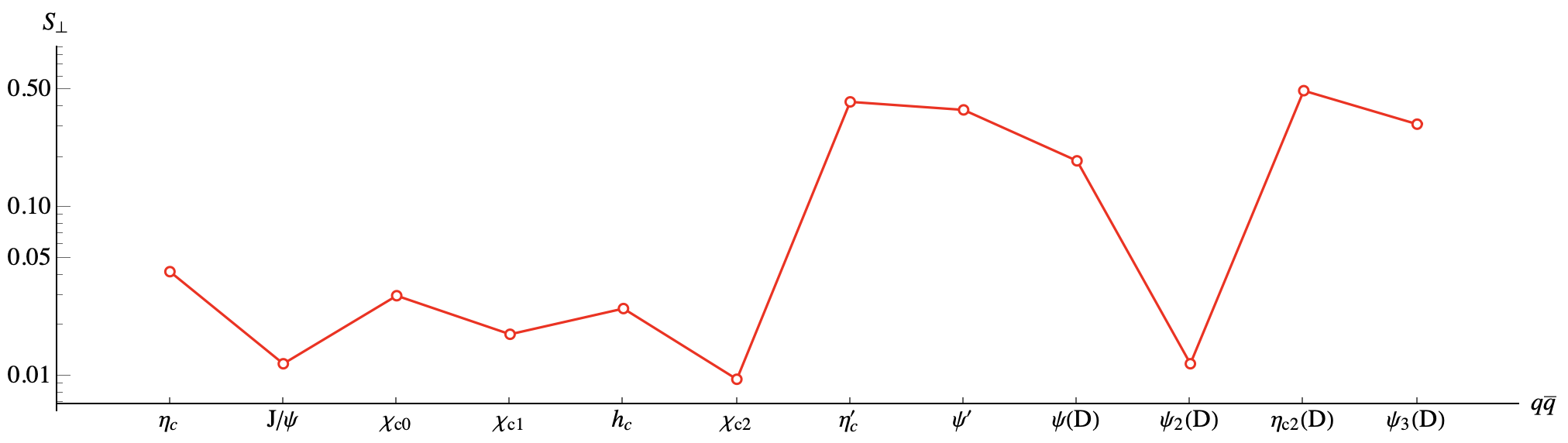}
\caption{Entanglement entropy between the transverse and the longitudinal d.o.f.'s for charmonium states obtained from BLFQ. The effective Hamiltonian adopts a separable confining (holographic confinement plus the LMZV longitudinal confinement) along with a non-separable one-gluon-exchange interaction \cite{Li:2017mlw}. 
}
\label{fig:EE}
\end{figure}

Of course, the true effective interaction between a quark and an antiquark may not be separable, for instance, the one-gluon-exchange interaction. A proper metric of the separability is the entanglement entropy $S_\perp$ between the transverse and the longitudinal d.o.f.'s, which is defined as the von Neumann entropy for the reduced density matrix, $S_\perp = -\mathrm{tr}\big[ \rho_\perp \log \rho_\perp \big]$ \cite{Eisert:2010}.  
Fig.~\ref{fig:EE} shows 
the entanglement entropy between the transverse and the longitudinal d.o.f.'s for charmonium states obtained from a non-separable interaction \cite{Li:2017mlw}. The effective Hamiltonian adopts a separable confining interaction (holographic confinement plus the Li-Maris-Zhao-Vary longitudinal confinement) along with a non-separable one-gluon-exchange interaction. As one can see, $S_\perp$ is generally small for the ground states -- indicating 
that these states are physically separable within this model. In this work, we adopt the separation of variable ansatz, which should provide a good description for the low-lying states. 

With this ansatz, the mass eigenvalues $M^2 = M_\perp^2 + M^2_\|$ and the wave function are also separable,
\begin{equation}
\psi(x, \vec k_\perp) = \sqrt{4\pi}\phi(\vec k_\perp/\sqrt{x(1-x})\chi(x).
\end{equation}
 We obtain two LFSWEs \cite{Chabysheva:2012fe}, 
\begin{align}
& \Big[{\vec\kappa^2_\perp} + V(\vec\zeta_\perp) \Big]\phi(\vec \kappa_\perp) = M^2_\perp \phi(\vec \kappa_\perp) \label{eqn:LFSWE_T}, \\
& \Big[\frac{m_q^2}{x}+\frac{m^2_{\bar q}}{1-x} + V_\|(\tilde z) \Big]\chi(x) = M^2_\| \chi(x). \label{eqn:LFSWE_L}
\end{align}
Here, the longitudinal effective potential $V_\|$ depends on the boost invariant longitudinal coordinate $\tilde z = \half p^+x^-$ (or other related longitudinal coordinate variables, e.g. (\ref{eqn:GT})), introduced by Miller and Brodsky \cite{Miller:2019ysh}. We note that even in the massless limit, Eq.~(\ref{eqn:LFSWE_L}) is still required to have non-trivial solutions. 
Hereafter we consider specific spin configurations (the leading-twist spin components) and the explicit spin indices will be suppressed, except for the purpose of state identification. 
N.B. the separation of variables ansatz requires that the variables $\vec \zeta_\perp$ and $\tilde z$ (or other related longitudinal coordinate variables) are independent. Since $\vec\zeta_\perp = \sqrt{x(1-x)}\vec r_\perp$ depends on $x$, $\tilde z$ should be defined as a differential operator against $\vec\zeta_\perp$, viz. $\tilde z = i\frac{\partial}{\partial x}\big|_{\vec \zeta_\perp}$.

It is convenient to normalize the transverse and longitudinal wave functions,
\begin{equation}\label{eqn:normalization}
 \int\frac{\dd^2 \kappa_\perp}{(2\pi)^2} \big| \phi(\vec\kappa_\perp) \big|^2 = 1, \quad
 \int_0^1\dd x \big| \chi(x) \big|^2 = 1.
\end{equation}
It is also useful to introduce a longitudinal wave function $X(x) = \sqrt{x(1-x)}\chi(x)$ with the normalization, 
\begin{equation}\label{eqn:normalization2}
 \int_0^1\frac{\dd x}{x(1-x)} \big| X(x) \big|^2 = 1.
\end{equation}
In our convention, $X(x)$ is proportional to the distribution amplitude. This is related to the alternative normalization convention of the wave function $\widetilde\Psi(x, \vec \zeta_\perp)$ or its transverse Fourier conjugate, 
\begin{equation}
\Psi(x, \vec \kappa_\perp) = \int \frac{\dd^2 \kappa_\perp}{(2\pi)^2} \widetilde\Psi(x, \vec \zeta_\perp) e^{-i\vec \kappa_\perp\cdot\vec \zeta_\perp}
= \psi(x, \vec k_\perp)/\sqrt{x(1-x)}
\end{equation}
Under this convention, the normalizations of the wave functions are (recall $\vec\kappa_\perp=\vec k_\perp/\sqrt{x(1-x)}$, $\vec \zeta_\perp = \sqrt{x(1-x)}\vec r_\perp$),
\begin{equation}
\int_0^1\dd x\int \frac{\dd^2k_\perp}{16\pi^3} \big| \Psi(x, \vec k_\perp/\sqrt{x(1-x)}) \big|^2 = 1, \quad 
\frac{1}{4\pi}\int_0^1 \frac{\dd x}{x(1-x)} \int \dd^2\zeta_\perp  \big| \widetilde\Psi(x, \vec r_\perp) \big|^2 = 1. 
\end{equation}
With this convention, the separable wave function takes the form, $\widetilde\Psi(x, \vec \zeta_\perp) = \sqrt{4\pi}\widetilde\phi(\zeta_\perp) X(x)$, 
where $\tilde\phi$ is the Fourier transform of $\phi$,
\begin{equation}
\tilde\phi(\zeta_\perp) = \int \frac{\dd^2 \kappa_\perp}{(2\pi)^2} e^{i\vec\kappa_\perp\cdot\vec\zeta_\perp} \phi(\vec\kappa_\perp).
\end{equation}
%
As an example, the pion wave function in LFH in the massless limit is $\chi(x) = 1$ and $X(x) = \sqrt{x(1-x)}$.

\subsection{Light-front holography}\label{sect:light-front_holography}

In LFH, the transverse effective potential is determined by the dilaton profile $\Phi(z)$ in AdS/QCD, $V_\perp = (1/2)\Phi'' + (1/4)\Phi'^2 + \big[(2J-3)/2z\big] \Phi'$ \cite{Brodsky:2008pg}. For the soft-wall profile, $V_\perp = \kappa^4 \zeta^2_\perp + 2\kappa^2(J-1)$. Here, $\kappa$ is the strength of the confinement in mass dimension. $J$ is the total angular momentum. The parton transverse separation $\zeta_\perp=\sqrt{x(1-x)} r_\perp$ is mapped to the fifth coordinate $z$ in Anti-de Sitter space.
The mass eigenvalues are, 
\begin{equation}
M^2_{nmJ} = 4\kappa^2 \Big(n +  \frac{J+|m|}{2}\Big).
\end{equation}
where $n, m$ are the radial and angular quantum numbers in the transverse plane ($\vec \zeta_\perp$). The corresponding wave functions are 2D harmonic oscillator functions,  
\begin{equation}
\phi_{nm}(\vec \zeta_\perp) =  \kappa\sqrt{\frac{n!}{\pi(n+|m|)!}} e^{-\frac{\kappa^2\zeta_\perp^2}{2}} (\kappa\zeta)^{|m|}_\perp L_n^{|m|}(\kappa^2 \zeta_\perp^2),
\end{equation}
where $n$ and $m$ are the transverse radial and angular quantum numbers. $N$ is a normalization constant. 
The ground state $n=0, m=0, J=0$ describes the pion, whose mass is predicted to vanish. Its wave function is Gaussian,  
\begin{equation}
\psi_\pi(x, \vec k_\perp) = \frac{4\pi}{\kappa}\exp\Big[-\frac{\vec k^2_\perp}{2\kappa^2 x(1-x)} \Big].
\end{equation}

To incorporate the finite quark masses, Brodsky and de Téramond employed a longitudinal wave function without explicitly introducing the longitudinal dynamics \cite{Brodsky:2008pf}. This ansatz, known as the invariant mass ansatz (IMA), is based on the observation that with the presence of finite quark mass, the light-front kinetic energy becomes,
\begin{equation}
\frac{\vec k^2_\perp}{x(1-x)} \; \to \; \frac{\vec k^2_\perp+m_q^2}{x} + \frac{\vec k^2_\perp+m_{\bar q}^2}{1-x}.
\end{equation}
Hence it is instructive to make the substitution in the wave functions. Then the pion wave function becomes, 
\begin{equation}
\psi_\pi(x, \vec k_\perp) = N \frac{4\pi}{\kappa} \exp\Big[-\frac{\vec k^2_\perp+m_q^2}{2\kappa^2 x(1-x)} \Big].
\end{equation}
In other words, the longitudinal wave function becomes,
\begin{equation}\label{eqn:IMA}
\chi_{\textsc{ima}}(x) = N \exp\Big[-\frac{m_q^2}{2\kappa^2 x(1-x)} \Big].
\end{equation}
The pion mass is also shifted to a non-zero value by,
\begin{equation}\label{eqn:Mpi_IMA}
M_\pi^2 = \int_0^1 \frac{\dd x}{2x(1-x)} \int \frac{\dd^2k_\perp}{(2\pi)^3} \big| \psi_\pi(x, \vec k_\perp) \big|^2 \frac{m_q^2}{x(1-x)}.
\end{equation} 
In the vicinity of the chiral limit, the theory predicts a near quadratic quark mass dependence of the pion mass, $M_\pi^2 \approx 2m_q^2 (\ln \kappa^2/m_q^2-\gamma_\textsc{e})$, where $\gamma_\textsc{e}\approx 0.577216$ is Euler's constant \cite{Li:2021jqb}. The scaling of the pion mass 
$M_\pi$ as a function of the quark mass $m_q$ is inconsistent with the results from chiral symmetry breaking, which predicts a linear dependence instead, a result also known as the Gell-Mann-Oakes-Renner (GMOR) relation \cite{Gell-Mann:1968hlm}, 
\begin{equation}
f^2_\pi M_\pi^2 =2 m_q\langle 0 | \bar q q | 0 \rangle + O(m_q^2).
\end{equation}
Gutsche et al. suggested replacing the longitudinal wave function (\ref{eqn:IMA}) from IMA by a power-law like function to generate the GMOR relation \cite{Gutsche:2012ez}. 

Another issue with IMA is the lack of longitudinal excitations, which leads to inconsistencies in state identification \cite{Li:2015zda, Li:2021jqb}. 
For example, in LFH the mass of $\rho(1D)$ is degenerate with the mass of $\rho(2S)$, and is 400 MeV lower than the mass of $\rho(1700)$, 
which is usually identified as $\rho(1D)$. Indeed, in LFH, $\rho(1700)$ is identified as $\rho(3S)$ instead. Similarly, $K_1(1400)$ in 
LFH is identified as $2{}^3P_1$ as opposed to the conventional identification $1{}^1P_1$. The resulting mass is higher than the experimental measurements. 

For states identified with quantum numbers $(n, m, S, J)$ in LFHQCD with IMA, $(-1)^JP = (-1)^{m+S+1}$, $C = (-1)^{m+S}$. From these relations, quantum number assignments of the scalars $a_0$ ($0^{++}$), the axial vectors $b_1$ ($1^{+-}$), and the axial tensors $a_2$ ($2^{++}$) etc.~are not compatible with the discrete symmetries.  The source of the discrepancy is the lack of longitudinal degrees of freedom (d.o.f.) in LFH with IMA. Indeed, in the basis function approach to the same quantum numbers, these states are excited in the longitudinal direction \cite{Li:2017mlw}. 

A possible alternative is to invoke the total orbital angular momentum $L$ and to assume that the LFH wave functions with IMA are only valid for states $L_z = |m| = L$. Note that $L$ is not a good quantum number in relativistic quantum mechanics. The obtained wave function may not be in leading twist ($L_z = 0$), either. 
By contrast, incorporating the longitudinal dynamics will provide the complete set of wave functions and the assignment of $P$ and $C$ does not rely upon these interpretations. 

\section{Longitudinal confinement}\label{sect:longitudinal_confinement}

The issues encountered in LFH point to the need for longitudinal dynamics. Indeed, as we mentioned, the decoupling of the transverse and longitudinal d.o.f. in Eqs.~(\ref{eqn:LFSWE_T}--\ref{eqn:LFSWE_L}) does not necessarily imply the absence of the longitudinal dynamics.  The presence of longitudinal confinement naturally leads to longitudinal excitations which ensure the correct state identification, similar to the non-relativistic cases but with exact discrete quantum numbers on the light front. Now, the new set of quantum numbers becomes $(n, m, l, S, J)$, where $l$ counts the number of nodes in the longitudinal direction. Under charge conjugation, $x \leftrightarrow (1-x)$ generates an extra sign $(-1)^l$. Hence, the charge conjugation quantum number should become $C = (-1)^{m+l+S}$. One can see that this resolves discrepancies associated with state identification in LFHQCD for states with longitudinal excitations. For non-separable states involving more than one eigenstate of the effective Hamiltonian, one can always go back Eqs.~(\ref{eqn:C}--\ref{eqn:mP}).  Hence, a longitudinal confinement is required to maintain the 3D structure of hadrons.

Of course, the longitudinal dynamics is also needed to incorporate the finite quark masses dynamically. 
This was first done by Chabysheva and Hiller utilizing the 't Hooft model \cite{Chabysheva:2012fe},
\begin{equation}
\big(V_\text{tH}\circ\chi\big) (x) = \frac{g^2}{\pi} \mathrm{P}\int \dd y \frac{\chi(x)-\chi(y)}{(x-y)^2},
\end{equation}
where the principle value prescription $\mathrm P ({1}/{x^2}) \equiv ({1}/{2})\big[{1}/{(x^2+i\epsilon)} + {1}/{(x^2-i\epsilon)}\big]$ is applied to the pole in the integrand. Chabysheva and Hiller focused on a specific solution with $m_q = g/\sqrt{\pi}$. The 't Hooft model as the longitudinal confinement was recently revisited by Ahmady et al.~\cite{Ahmady:2021lsh, Ahmady:2021yzh}.

 Another longitudinal confinement based on a construction of the harmonic oscillator potential was proposed by G\l{}azek and Trawi\'nski (GT) \cite{Glazek:2013jba, Trawinski:2014msa}. They introduce a third momentum $\kappa_\| \equiv (m_q+m_{\bar q})(x-\hat m_q)/\sqrt{x(1-x)}$, where $\hat m_q = m_q/(m_q+m_{\bar q})$. The longitudinal confining potential is defined as 
the harmonic oscillator potential of the conjugate coordinate $\zeta_\|$ of $\kappa_\|$, viz.
\begin{equation}\label{eqn:GT}
V_\text{GT} = \kappa^4 \zeta_\|^2.
\end{equation}
The resulting mass squared eigenvalues obey the linear Regge trajectory, $M^2_l = \kappa^2 (l+\frac{1}{2})$. 
The ground-state wave function is Gaussian,
 \begin{equation}
 \chi_\text{GT}(x) = N \exp\Big[-(m_q+m_{\bar q})^2\frac{(x-\hat m_q)^2}{2\kappa^2 x(1-x)}\Big]
 = N' \exp\Big[-\frac{1}{2\kappa^2}\Big(\frac{m_q^2}{x} + \frac{m_{\bar q}^2}{1-x}\Big)\Big]
 \end{equation}
 Therefore, this model can be viewed as a generalization of the IMA. 

The quark masses adopted in Refs.~\cite{Chabysheva:2012fe, Glazek:2013jba, Trawinski:2014msa} are of the order of the constituent quark mass ($\sim$ 300 MeV). In Ref.~\cite{Li:2015zda}, Li, Maris, Zhao and Vary proposed an analytically solvable longitudinal confining potential and applied it to the quarkonium in the framework of BLFQ. 
In Ref.~\cite{Li:2021jqb}, we further show that in the vicinity of the chiral limit, the LMZV model reproduces the GMOR relation as well as the signature power-law wave function. This is sometimes termed BLFQ${}_0$ to distinguish from the applications to the 
heavy flavors where the longitudinal coupling $\sigma$ is chosen to match to the transverse confining strength in the 
NR limit. 
The longitudinal confinement takes the form,
\begin{equation}
\big(V_\text{LMZV}\circ\chi\big)(x) = -\sigma^2\frac{\dd}{\dd x}\big(x(1-x)\frac{\dd}{\dd x}\chi(x)\big),
\end{equation}
where $\sigma$ is the strength of the longitudinal confinement. 
The same longitudinal confinement is adopted by de Téramond and Brodsky in Ref.~\cite{DeTeramond:2021jnn} to incorporate chiral symmetry breaking. Both works note the close relation between the LMZV/BLFQ${}_0$ model and the 't Hooft model. The latter is known to exhibit the chiral symmetry breaking via the Berezinskii–Kosterlitz–Thouless (BKT) mechanism \cite{Ji:2020bby}. 
From the ground state masses in the vicinity of the chiral limit $M^2_\textsc{lmzv} = \sigma (m_q+m_{\bar q}) + O(m_{q,\bar q}^2)$, and $M^2_\text{tH} = g\sqrt{\frac{\pi}{3}} (m_q+m_{\bar q}) + O(m_{q,\bar q}^2)$,  we can identify $\sigma =  g\sqrt{\frac{\pi}{3}} = |\langle \bar q q \rangle|/f^2_\pi$.  The longitudinal wave function of the LMZV/BLFQ${}_0$ model is,
\begin{equation}
\chi_\textsc{lmzv}(x) = N x^{\alpha_1}(1-x)^{\alpha_2}
\end{equation}
where $\alpha_i = m_i/\sigma$. The 't Hooft wave function is, 
\begin{equation}
\chi_\text{tH}(x) \dot= N x^{\beta_1}(1-x)^{\beta_2}.
\end{equation}
The exponents $\beta_{1,2}$ satisfy $\pi m_i^2/g^2 = 1 - \pi\beta_i\cot \pi\beta_i$. The solution in the vicinity of the chiral limit is, $\beta_i = (m_i/g)(\sqrt{3/\pi}) = \alpha_i$, identical to those of the LMZV/BLFQ${}_0$ model. In other words, the ground-state mass eigenvalue and eigenfunctions of the both models are identical. These two models differ in the excited state masses. For large excitation, the mass eigenvalues of the 't Hooft model obey the linear Regge trajectory, $M^2_n \propto n$, whereas those of the LMZV/BLFQ${}_0$ model are quadratic, $M^2_n \propto n(n+1)$. 
A detailed comparison of the hadron spectra using the 't Hooft model and the LMZV/BLFQ${}_0$ model with the original predictions from LFH with IMA is reported by Ahmady et al.~\cite{Ahmady:2021yzh}. 

Note that Ahmady et al. imposed a constraint $l \ge n + m$ on the hadron spectra and abandoned low-lying states without the required longitudinal excitations. For example, the scalar ($0^{++}$) $a_0(980)$ is identified as $(n, m, S) = (0, 1, 1)$ in LFH with IMA, with a predicted mass $M_\textsc{ima} = 0.76\,\text{GeV}$. In the LMZV/BLFQ${}_0$ model, this particle is identified as $(n, m, l, S) = (0, 0, 1, 1)$ with a predicted mass $M_\textsc{lmzv} = 0.91\,\text{GeV}$. Both models are consistent with the quark model identification $1{}^3\!P_0$ while the latter improves the theoretical prediction. By contrast, in Ahmady et al., this state is identified as a tetra-quark with $(n_T, m_T, l_T, S_T) = (0, 0, 7, 0)$. The resulting mass is $M_\textsc{amkms} = 1.39\,\text{GeV}$.  In fact, in Ref.~\cite{Ahmady:2021yzh}, all scalars, including $\chi_{c0}(1P)$ and $\chi_{b0}(1P)$, are identified as tetra-quarks.

The pion wave functions of the LMZV/BLFQ${}_0$ model are similar to those adopted by Gutsche et al. \cite{Gutsche:2012ez}. The pion mass  is exactly two times of  Gutsche et al.'s result. The difference is caused by the presence of the confining interaction in LMZV/BLFQ${}_0$ model, which contributes to the other half of the pion mass.

\begin{figure}
\centering
\includegraphics[width=0.65\textwidth]{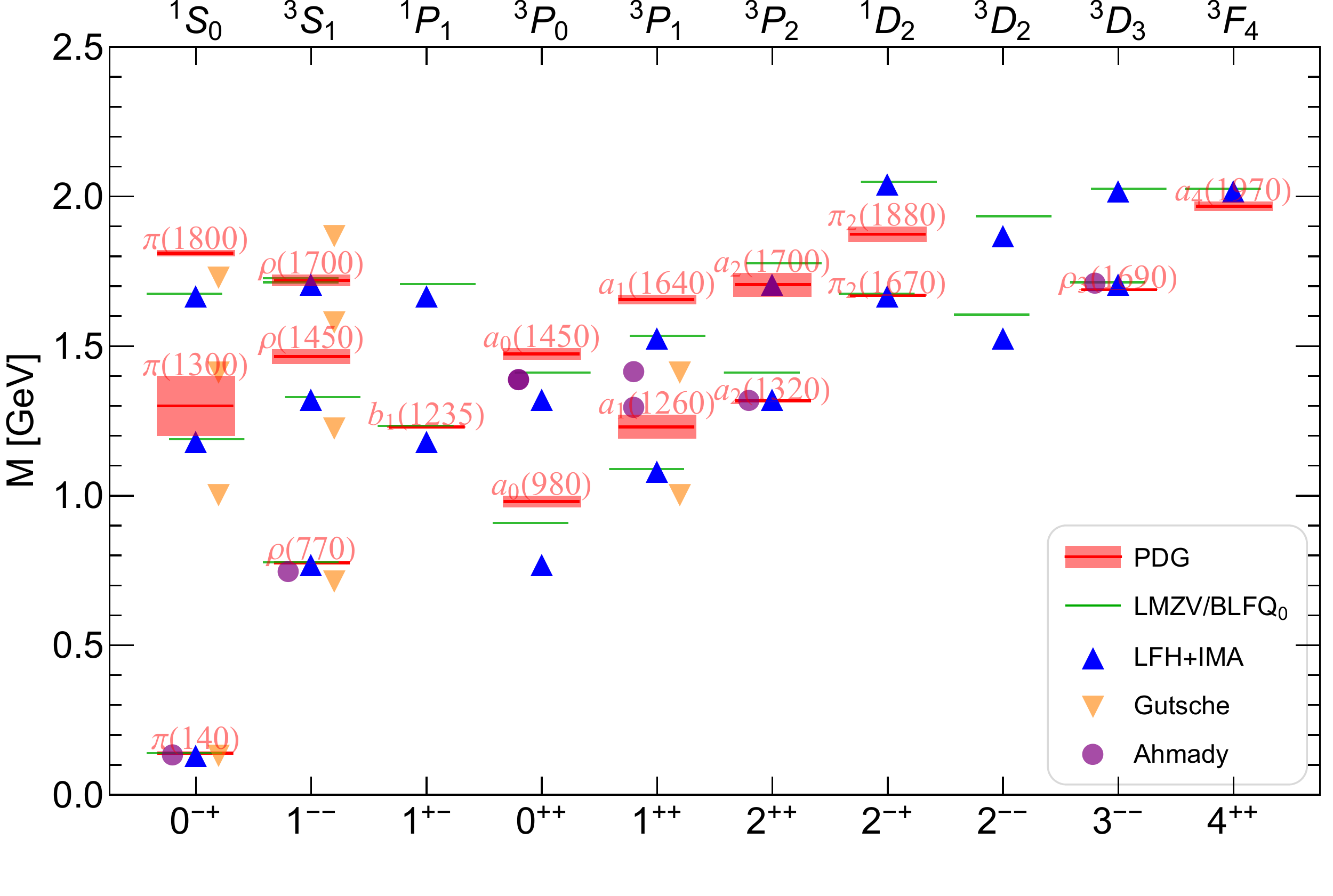}

\includegraphics[width=0.65\textwidth]{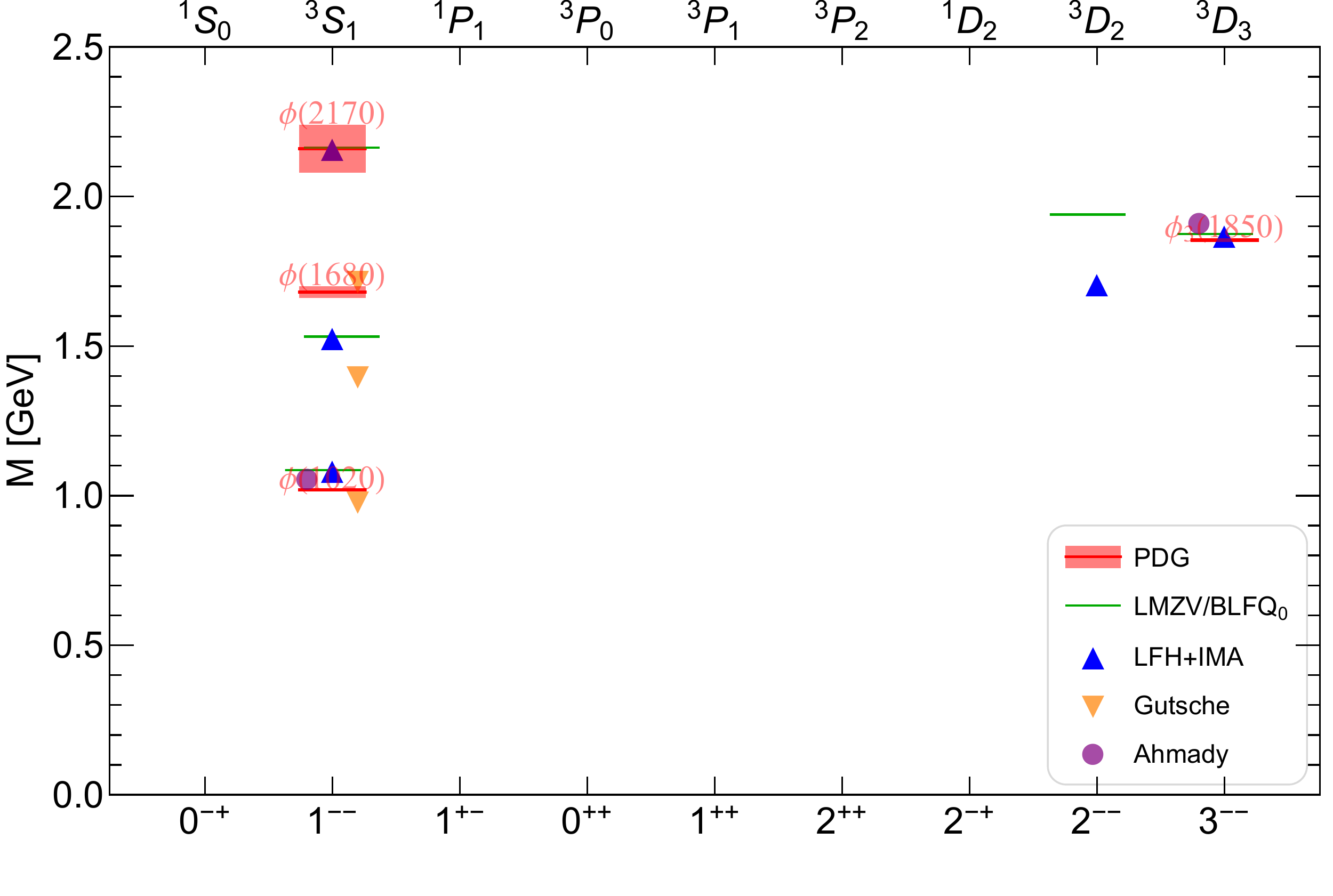}
 
\includegraphics[width=0.65\textwidth]{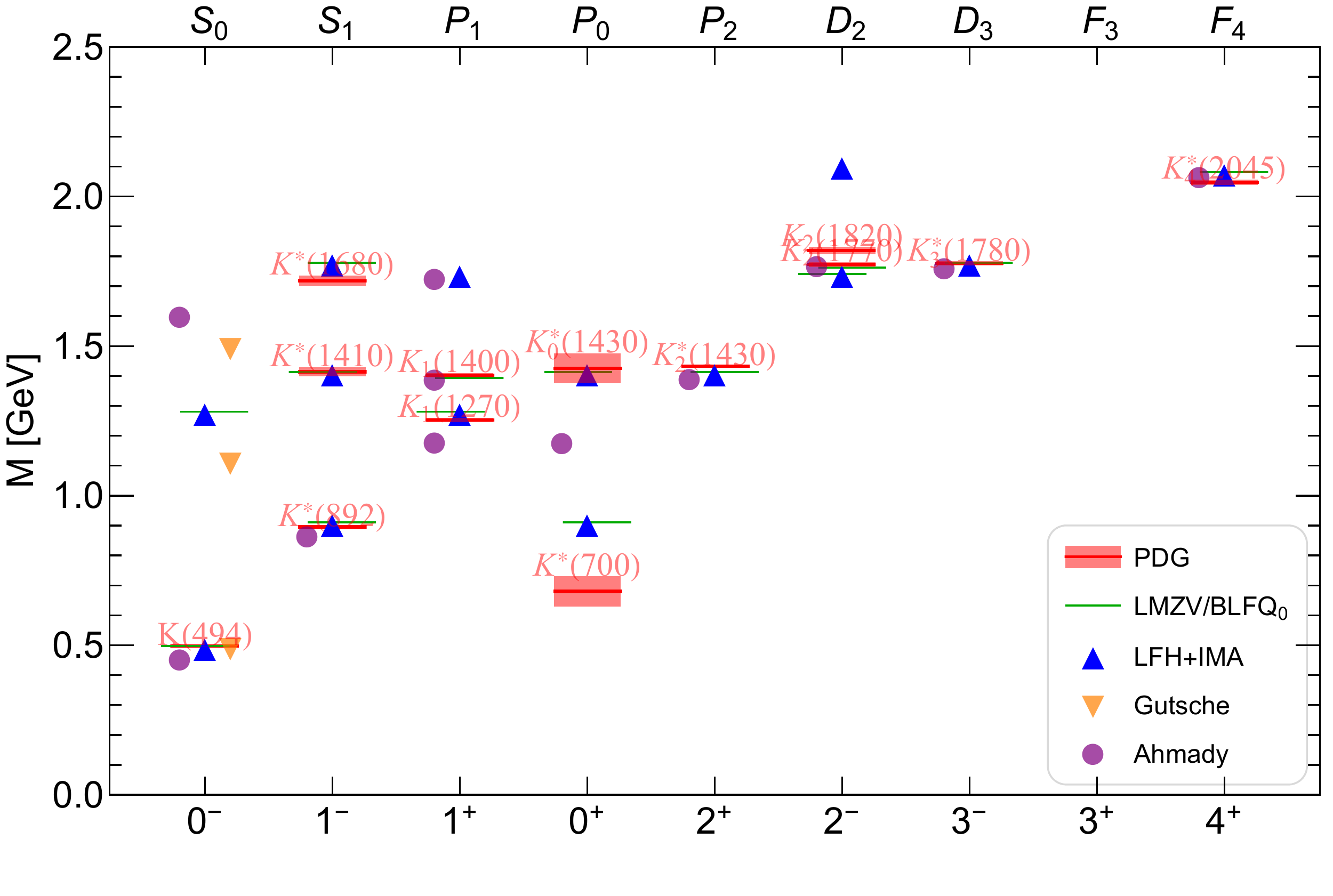}
\caption{Comparison of unflavored light meson spectra and kaons as predicted by LFH with IMA \cite{Brodsky:2014yha}, with LMZV/BLFQ${}_0$ \cite{Li:2021jqb} and with 't Hooft potential as implemented by Ahmady et al. \cite{Ahmady:2021yzh}.  Results from Gutsche et al. which adopt LFH with a longitudinal wave function similar to LMZV are also attached for comparison \cite{Gutsche:2012ez}. }
\label{fig:spectra}
\end{figure}

Figure~\ref{fig:spectra} compares the mass spectra of unflavored light mesons and kaons as predicted by LFH with IMA \cite{Brodsky:2014yha}, LFH with with the power-law like wave function by Gutsche et al.\cite{Gutsche:2012ez}, LFH with LMZV/BLFQ${}_0$ \cite{Li:2021jqb} and LFH with 't Hooft potential by Ahmady et al. \cite{Ahmady:2021yzh}. Overall, LMZV/BLFQ${}_0$ provides significant improvement of LFH with IMA towards the experimental data.

In the non-relativistic (NR) limit, rotational symmetry requires that the combination of the transverse and longitudinal confining interactions is a rotational invariant. The transverse confinement reduces to a harmonic oscillator potential in this limit $\kappa^4 \zeta_\perp^2 \to ({\kappa^4}/{4})\vec r_\perp^2$. It is expected that the longitudinal confinement also reduces to the quadratic form. The GT model satisfies rotational symmetry in the NR limit by construction. The LMZV model also satisfies this symmetry with the identification $\sigma = \kappa^2/(m_q+m_{\bar q})$ in the NR limit. It is curious to note that heavy quark effective theory requires $\kappa \propto \sqrt{M_H} \approx \sqrt{m_q + m_{\bar q}}$. Hence, the confining strength 
$\sigma$ approaches to a constant in the heavy quark limit. From the fits to the physical quark masses, $\sigma_\textsc{hq} \approx 0.24\,\text{GeV}$, where "HQ" stands for "Heavy Quark" \cite{DeTeramond:2021jnn}. 
Physically, this value is determined by the gluon condensate $\langle G_{\mu\nu} G^{\mu\nu} \rangle$, which is in general different from the values in the light sector $\sigma = \langle \bar q q\rangle /f_\pi^2 \approx 0.6\,\text{GeV}$ determined by the quark condensate \cite{DeTeramond:2021jnn}. 

In coordinate space, the 't Hooft interaction is (recall $\tilde z = \frac{1}{2}P^+x^-$),
\begin{equation}
V_\text{tH} = g^2 \big|\tilde z\big|
\end{equation}
In the NR limit, $\tilde z \to -2M r_z$.  The rotational symmetry is not restored with the quadratic transverse confinement from LFH. Note that in Ref.~\cite{Ahmady:2021yzh}, Ahmady et al. argued that rotational symmetry can be restored with the matching $g=\kappa$, as the transverse LFH confinement and the 't Hooft interaction are equivalent to the c.m. instant form potential $V_\perp = (\kappa^2/2)r_\perp$ and $V_\| = (g^2/2)r_z$, respectively, based on an relation $U_\text{LF} = V_\text{IF} + 2(m_q+m_{\bar q})V_\text{IF}$ proposed by Trawi\'nski et al. \cite{Trawinski:2014msa}. Even if this relation is held in the NR limit, the corresponding potential $V = V_\perp + V_\| = (\kappa^2/2)(\sqrt{r_x^2+r_y^2}+r_z)$ is still not rotationally invariant. Shuryak and Zahed proposed an alternative confining interaction that generalizes the 't~Hooft model to 3+1D while combining the light-front holographic interaction \cite{Shuryak:2021hng, Shuryak:2021mlh}.
\begin{equation}
V_\text{SZ} = 2\sigma_T \sqrt{\tilde z^2 + M^2 r^2_\perp}.
\end{equation}
The rotational symmetry is restored in the NR limit in this model.

\begin{figure}
\centering 
\includegraphics[width=0.45\textwidth]{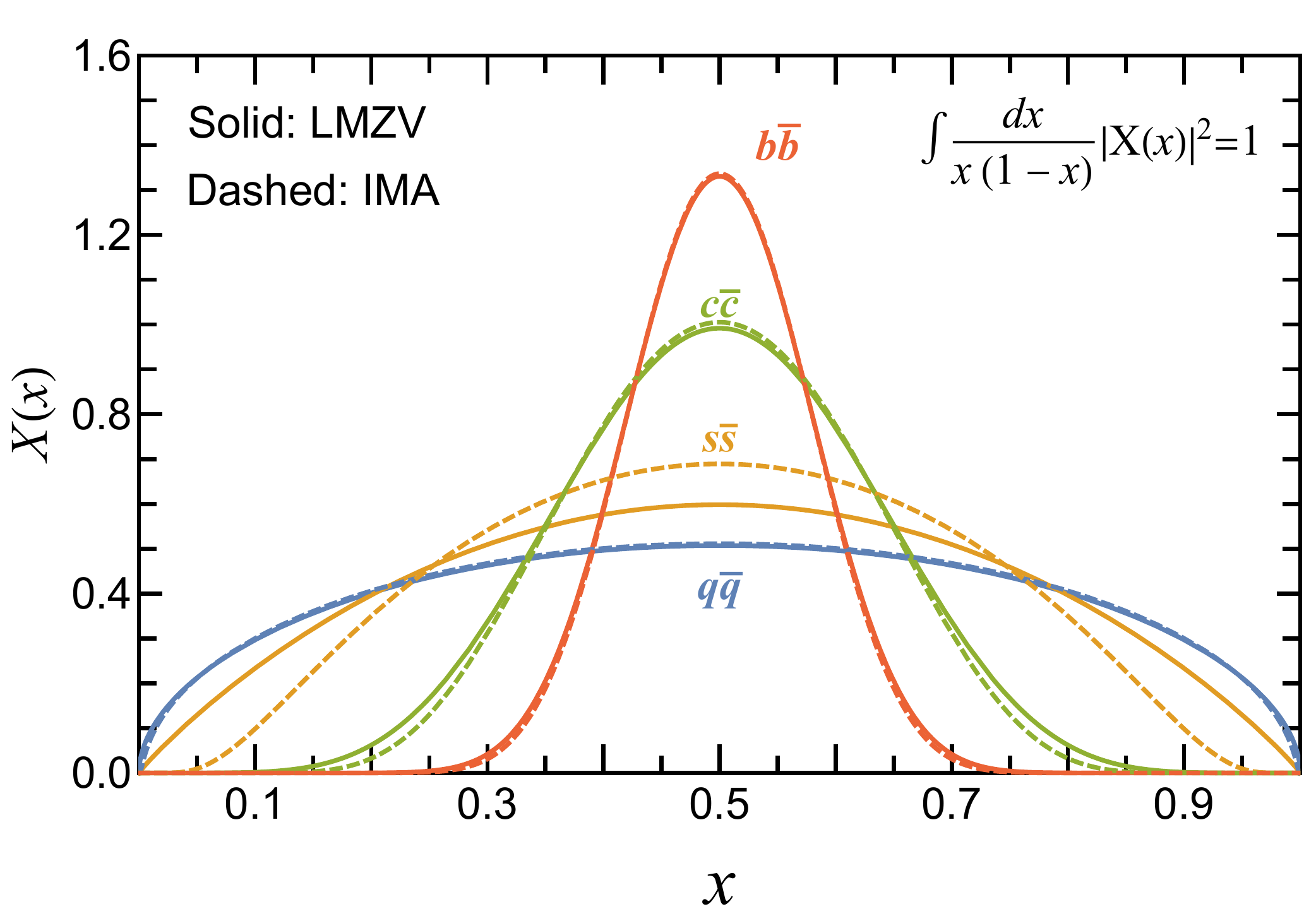}
\includegraphics[width=0.45\textwidth]{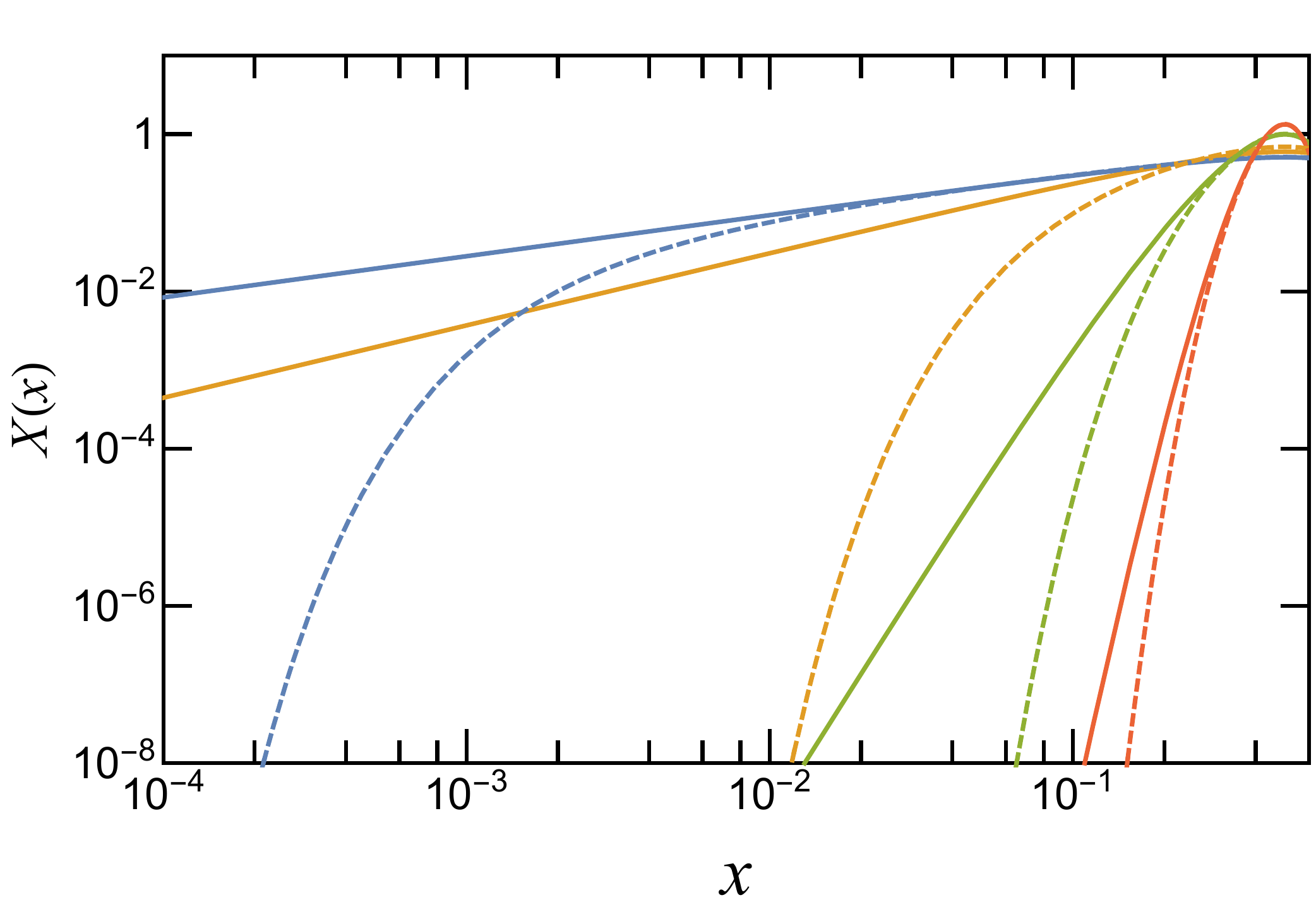}
\caption{Comparison of the longitudinal wave functions of ground states from the IMA and from the LMZV model. 
}
\label{fig:NRreduction}
\end{figure}

Figure~\ref{fig:NRreduction} compares the ground-state longitudinal wave functions $X(x) = \sqrt{x(1-x)}\chi(x)$ from the IMA and from the LMZV model. The parameters of the IMA are adopted from LFHQCD \cite{Brodsky:2014yha, Nielsen:2018ytt}. For $q\bar q$ and $s\bar s$, the parameters of the LMZV model are adopted from Ref.~\cite{Li:2021jqb}. For $c\bar c, b\bar b$, the parameters are matched to the IMA parameters with the NR reduction, $\sigma = \kappa^2/(m_q+m_{\bar q})$.
As we can see, the main difference between these two confining potentials is the endpoint asymptotics. 
Despite the striking similarities of the longitudinal wave functions from IMA and the LMZV model shown in Fig.~\ref{fig:NRreduction}, we would like to emphasize that the endpoint behaviors are dramatically different and this fact has important physical consequences. 
In the 't Hooft and LMZV models, the endpoint asymptotics is a direct consequence of chiral symmetry breaking and the exponent is related to the chiral condensate. The endpoint behavior also has a dramatic impact on hadronic observables in high energy collisions as hard kernels $T_H$ are sensitive to the endpoint singularities \cite{Lepage:1980fj}. 

\begin{figure}
\centering 
\includegraphics[width=0.5\textwidth]{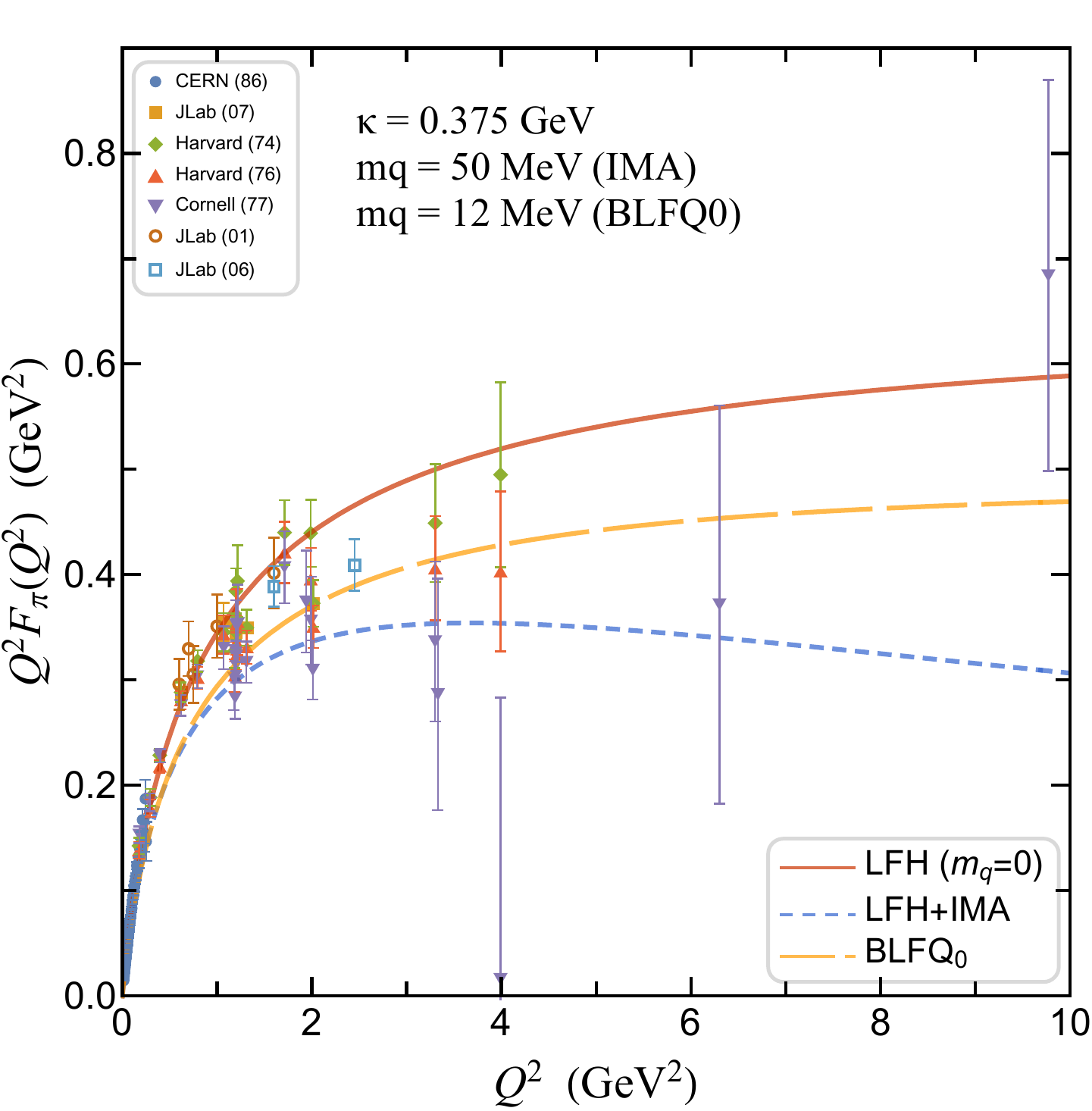}
\caption{Large-$Q^2$ asymptotics of the pion form factor $Q^2F_{\pi}(Q^2)$ from LFH (solid), LFH with IMA (dotted), and LMZV/BLFQ${}_0$ (dashed), for the original parameters of LFH with $\kappa = 0.375\,\mathrm{GeV}$. The quark masses are $m_{\{u,d\}} = 50\,\mathrm{MeV}$ for LFH with IMA and $m_{\{u,d\}} = 12\,\mathrm{MeV}$ for LMZV/BLFQ${}_0$.}
\label{fig:pion_FF}
\end{figure}

As an example, one may consider the pion form factor \cite{Li:2021jqb}. In light-front holography, the pion form factor is \cite{Brodsky:2007hb, Brodsky:2014yha}, 
\begin{equation}
F_\pi(Q^2) = \int_0^1 \frac{\dd x}{x(1-x)} \big| X(x) \big|^2 \exp\Big(-\frac{1-x}{x} \frac{Q^2}{4\kappa^2}\Big).
\end{equation}
The large $Q^2$ behavior of $F_\pi(Q^2)\sim 1/Q^{1+2\delta}$ is directly related to the endpoint asymptotics of the (normalized) distribution amplitude 
$X \sim (1-x)^\delta$ \cite{Lepage:1980fj}. In 't Hooft and LMZV models, $\delta = 1/2 + m_q/\sigma$. Hence,  $Q^2 F_\pi(Q^2) \sim 1/\log Q^2$ in the vicinity of the chiral limit. This is in contrast to the LFH prediction $Q^2F_{\pi}(Q^2) \to \text{const}$ \cite{Brodsky:2007hb} and to the prediction from LFH with IMA, $Q^2F_{\pi}(Q^2) \to e^{-cQ^2}$. These results are shown in Fig.~(\ref{fig:pion_FF}), where we adopt parameters from the original LFH \cite{Brodsky:2007hb}. Of course, a comprehensive investigation of the pion form factor in holography requires the dressed holographic current \cite{Brodsky:2014yha}, which is beyond the scope of the present work. We only point out that the confined current has similar dependence on the endpoint behavior with the bare current at large $Q^2$ \cite{Brodsky:2011xx}.

A similar example is the pion radiative transition form factor\footnote{Here we have neglected the evolution of the pion distribution amplitude since the evolution is extremely slow for large but finite $Q^2$, say $1\;\text{GeV}^2 \lesssim Q^2 \lesssim 1000\;\text{GeV}^2$.} \cite{Brodsky:2011xx,Swarnkar:2015osa}, 
\begin{equation}
Q^2F_{\pi\gamma}(Q^2) = \frac{4}{\sqrt{3}} \int_0^1 \dd x \frac{\phi_{\pi}(x, \tilde Q)}{1-x}. \quad (\tilde Q = (1-x)Q)
\end{equation}
Its value at large $Q^2$ is predicted to approach different asymptotic values with different asymptotics at the endpoints.

\section{Sturm-Liouville Theory}\label{sect:sturm-liouville_theory}

After a comparison of various models, a natural question to ask is whether there is freedom for further proposals. This task is not as easy as it seems. Because the construction of a valid Hermitian operator in the functional space may be jeopardized by the singularities in the light-front kinetic energy term. Weller \& Miller unified the previous longitudinal potentials with Miller and Brodsky's spatial coordinate $\tilde z$ \cite{Weller:2021wog}. An alternative approach is the momentum representation, 
\begin{equation}
(V_\text{F}\circ\chi)(x) = \int_0^1 \dd y K(x, y) \chi(y)
\end{equation}
The kernel $K(x, y)$ may be related to the coordinate space ($\tilde z$) potential.
\begin{equation}
K(x-x') = \int \dd \tilde z \, e^{ix\tilde z} V_\|(\tilde z).
\end{equation} 

The 't Hooft model $K(x-x') \propto 1/(x-x')^2$ corresponds to the linear potential $V_\text{tH} = |\tilde z|$, as mentioned. In general, the power-law potential $|\tilde z|^p$ leads to a kernel $K(x-x') = \frac{1}{|x-x'|^{p+1}}$ ($p \notin 2\mathbb N$).
The kernel for the quadratic potential $\tilde z^2$ is the derivative of a Dirac-$\delta$, which is more convenient to be expressed as differential operators. Due to the low dimensionality, these constructions may be associated with severe singularities.

Another useful tool is the Sturm-Liouville (SL) theory. The corresponding interaction takes the form,
\begin{equation}
V_\text{SL} = - \partial_x p(x) \partial_x + s(x). 
\end{equation}
Clearly, the LMZV interaction belongs to this category. 
The longitudinal Schr\"odinger equation becomes  (\ref{eqn:LFSWE_L}), 
\begin{equation}
\Big[\frac{m_q^2}{x}+\frac{m^2_{\bar q}}{1-x} + V_\text{SL} \Big]\chi(x) = M^2_\| \chi(x). 
\end{equation}
It may be written as a Sturm-Liouville problem
\begin{equation}\label{eqn:SLEVP}
-\big( p(x) \chi'(x) \big)' + q(x)\chi(x) = \lambda \chi(x), \quad (\chi(0) = \chi(1) = 0)
\end{equation} 
where $q(x) = m_q^2/x + m^2_{\bar q}/(1-x) + s(x)$.  
Note that the derivative $\partial_x$ is taken with respect to $\vec\zeta_\perp$ not $\vec r_\perp$, viz $\partial_x = (\partial/\partial x)_{\vec\zeta_\perp}$. In the NR limit, rotational symmetry can be restored if $p(x=\hat m_q) = \kappa^4/[4(m_q+m_{\bar q})^2]$.

We shall call this class of potentials as the Sturm-Liouville (SL) type, since the longitudinal LFSWE forms a Sturm-Liouville eigenvalue problem (SLEVP). The SL longitudinal confinement is naturally compatible with the transverse holographic confinement in the NR limit. The advantage of these extensions is that the known properties of the SLEVP, e.g.  the distribution of eigenvalues,
 can be applied to the longitudinal LFSWE.

To the first approximation, function $p(x)$, stemming from the inter-quark potential, should be symmetric wrt the interchange $x \leftrightarrow 1-x$. Therefore, it can be written as a function of $x(1-x)$ and $(x-\frac{1}{2})^2$. Unless the quark masses are equal, $(x-\frac{1}{2})^2$ is not consistent with the NR reduction $x \approx \hat m_q$ and $x \approx \hat m_{\bar q}$. Therefore, the natural choice of $p(x)$ is $p(x) = \sigma^2 [x(1-x)]^\gamma$. A typical source of $s(x)$ is the self-energy correction. A dynamically generated mass may have the asymptotics $\sim B/x(1-x)$, which can be absorbed into the quark mass. In general, we can assume that $s(x)$ is no more singular than the light-front kinetic energy. A convenient choice is $s(x) = 0$. 

A comment on the normalization is in order. If, instead, the following normalization is adopted, 
\begin{equation}
\int_0^1\frac{\dd x}{x(1-x)} \big| X(x) \big|^2 = 1,
\end{equation}
where a non-trivial weight function $w(x) = 1/[x(1-x)]$ is introduced, the Sturm-Liouville problem for $X(x)$ should be, 
\begin{equation}
- \big( P(x) X'(x) \big)' + Q(x)X(x) = \lambda w(x) X(x), \quad (X(0) = X(1) = 0).
\end{equation} 
where 
\begin{align}
& P(x) = p(x) w(x) = \frac{p(x)}{x(1-x)}, \\
& Q(x) = q(x)w(x) - \frac{pw'^2}{4w} + \frac{1}{2}(pw')' = \frac{q(x)}{x(1-x)} - \frac{8x^2-8x+3}{4x^3(1-x)^3}p(x) + \frac{2x-1}{2x^2(1-x)^2}p'(x). 
\end{align}
For example, the corresponding interaction for the LMZV model reads,
\begin{equation}
U_\| = -\sigma^2 \partial^2_x + \frac{\sigma^2}{4x^2(1-x)^2}
\end{equation}

Now let us turn to the general properties of the SLEVP, which is covered by the Sturm-Liouville theorem \cite{Krall:1986, Zettl:2005}. The SL problem (\ref{eqn:SLEVP}) is called singular if either $p(x) = 0$ at the endpoints or $q(x)$ has singularities. Otherwise, the SL problem is called regular. Clearly, the light-front kinetic energy term in $q(x)$ leads to a singular SLEVP. We remind that the Sturm-Liouville theorem is only applicable to the regular SLEVP or the singular SLEVP in limit circle non-oscillation (LCNO) \cite{Krall:1986, Zettl:2005}. For other singular SLEVPs, including the limit circle oscillation (LCO) case and the limit point (LP) cases, the properties of the eigenvalues and eigenfunctions are not known in general. For example, the eigenvalues may not be bounded from below; the eigen-solutions may have infinite oscillations near the endpoints. Such pathological solutions are known in finite truncations of relativistic bound-state equations, such as Bethe-Salpeter equations. 
What is worse, the classification of singular SLEVP generally relies on the solution. Mathematically, only a finite number of cases are classified. A catalog of the singular SLEVP is given by Ref.~\cite{Everitt:2000}. 

In fact, the most famous SLEVP of the LCNO case on the finite interval $[0,\, 1]$, is the familiar Legendre function, corresponding to $p(x) = x(1-x)$, and $q(x) = 0$. Here, we have applied a shift $x = (t+1)/2$. The eigenvalues $\lambda_\ell = \ell(\ell+1)$ and the eigen-functions are the Legendre polynomials $P_\ell(2x-1)$. The longitudinal confinement $V_\| = -\sigma^2\partial_x \big(x(1-x)\partial_x\big)$ covers a number of well-known special functions, the Legendre function, the Chebyshev function and the Gegenbauer function. The most general case is the Jacobi function which is applicable to the general unequal quark mass case,
\begin{equation}
\Big[\frac{m_q^2}{x}+\frac{m^2_{\bar q}}{1-x} - \sigma^2\frac{\dd}{\dd x} x(1-x) \frac{\dd}{\dd x}  \Big]\chi(x) = M^2_\| \chi(x). 
\end{equation}
The associated SL problem is,
\begin{equation}\label{eqn:LFSE1D}
-\big( x(1-x)\chi'\big)' + \Big(\frac{\beta^2}{4x} +  \frac{\alpha^2}{4(1-x)}\Big) \chi = \lambda \chi, \quad (\chi(0) = \chi(1) = 0)
\end{equation} 
where $\alpha = 2m_{\bar q}/\sigma, \beta = 2m_{q}/\sigma$. The solution is the Jacobi polynomials: 
\begin{equation}
\chi_n(x) = N_n x^{\frac{\beta}{2}}(1-x)^{\frac{\alpha}{2}}P_{n}^{(\alpha,\beta)}(2x-1).
\end{equation}
The eigenvalues are $\lambda_n = (n+\alpha+\beta) (n+1+\alpha+\beta)\propto n^2$. The mass eigenvalue $M^2_{\|n}= (m_q+m_{\bar q})^2 +  \sigma (m_q+m_{\bar q}) (2n+1) + \sigma^2n(n+1)$. 
The ground state mass obeys the GMOR relation, $M^2 = (m_q+m_{\bar q})^2 + \sigma (m_q+m_{\bar q})$. The ground state wave function is power-law like.

The Jacobi function belongs to an even broader class of special functions, the hypergeometric function \cite{AS, DLMF}. It is expected that a broader class of longitudinal confining potentials may be constructed there with hypergeometric functions as the solutions. Unfortunately, the classification of the general 
hypergeometric differential equation on the interval $[-1, 1]$ is not known. 

A similar situation occurs for another large class of second order differential equations on $[0, 1]$, the Heun equation \cite{Everitt:2000, Slavyanov:2000}. There are at least two special cases of the Heun equation that are related to the LMZV interaction. The first one stems from the azimuthal part of the three-body Schrödinger equation that describes the hydrogen-molecule ion $H^+_2$ (two protons plus one electron) \cite{Slavyanov:2000}. It corresponds to a longitudinal potential $V_\| = -\sigma^2 \big[
\partial_x \big(x(1-x)\partial_x\big) + \eta (2x-1)^2\big]$. The eigensolutions are shown to acquire additional exponential factors: $\exp(\pm c \big|2x-1\big|)$ \cite{Mitin:2015}. The second case is the Teukolsky equation that describes the stability of the Kerr black hole. It corresponds to a longitudinal potential $V_\| = -\sigma^2 \big[\partial_x \big(x(1-x)\partial_x\big) + a^2 (2x-1)^2 + 4a (2x-1) \big]$ and a mass function $M = m+b(2x-1)$ \cite{Slavyanov:2000}.  The mass correction can also be absorbed into the potential. 

The scaling of the eigenvalues is given by the Atkinson–Mingarelli theorem \cite{Atkinson:1987}. For large $n$, $\lambda_n$ scales quadratically, 
\begin{equation}\label{eqn:Atkinson}
\lambda_n \sim \frac{n^2\pi^2}{\Big(\int_a^b \frac{1}{\sqrt{p(x)}}\Big)^2}
\end{equation}
From this theorem, the eigenvalues of the SL confining potentials are generally quadratic, unless the integral in the denominator diverges. 

A natural generalization of the LMZV model is to consider $p(x) = \sigma^2 x^\gamma (1-x)^\gamma$. Except for $\gamma=1$, the obtained SLEVPs are not well studied in mathematics.  
However, as we will see, at least some of these problems (e.g., $\gamma=0$) are physically well defined. We will investigate the longitudinal potentials with $\gamma=0,1,2,3$. The results are summarized in Table~\ref{tab:summary}. Note that for $\gamma \ge 2, \gamma \in \mathbb Z$, the estimate Eq.~(\ref{eqn:Atkinson})
is not applicable and other distributions of the eigenvalues are possible. 

\subsection{Case I: $p(x)=\sigma^2$} 

In the first case, we take $p(x)=\sigma^2, s(x)=0$. This is in fact one of the simplest cases. The longitudinal confining potential $V_\| = \sigma^2 \tilde z^2$ is quadratic in the longitudinal coordinate space. 
Then the longitudinal LFSWE becomes, 
\begin{equation}
\Big[\frac{m_q^2}{x}+\frac{m^2_{\bar q}}{1-x} - \sigma^2\frac{\dd^2}{\dd x^2}  \Big]\chi(x) = M^2_\| \chi(x). 
\end{equation}
The associated Sturm-Liouville problem, for the equal mass case $m_q = m_{\bar q}$, is,
\begin{equation}\label{eqn:LFSE1D}
-\chi''(x) + \frac{\mu^2}{x(1-x)} \chi(x) = \lambda \chi(x), \quad (\chi(0) = \chi(1) = 0)
\end{equation} 
where $\mu = m_q/\sigma$, and $\lambda = M^2/\sigma^2$.

This is a 1D Schr\"odinger equation with a potential $V \propto 1/(x(1-x))$. In the chiral limit ($\mu=0$), the potential well becomes 
an infinite square potential well. The solution in this case is the trigonometry function $\chi_n(x) = \sqrt{2}\sin n\pi x$, and the $n$-th eigenvalue $\lambda_n = n^2\pi^2$, ($n=1,2,\cdots$).\footnote{The massless solution $\lambda=0$, $\chi(x) = a + b x$, is eliminated by the boundary condition.} The ground state energy is the zero-point energy, $M^2 = \sigma^2\pi^2$.

\begin{figure}
\centering
\includegraphics[width=0.4\textwidth]{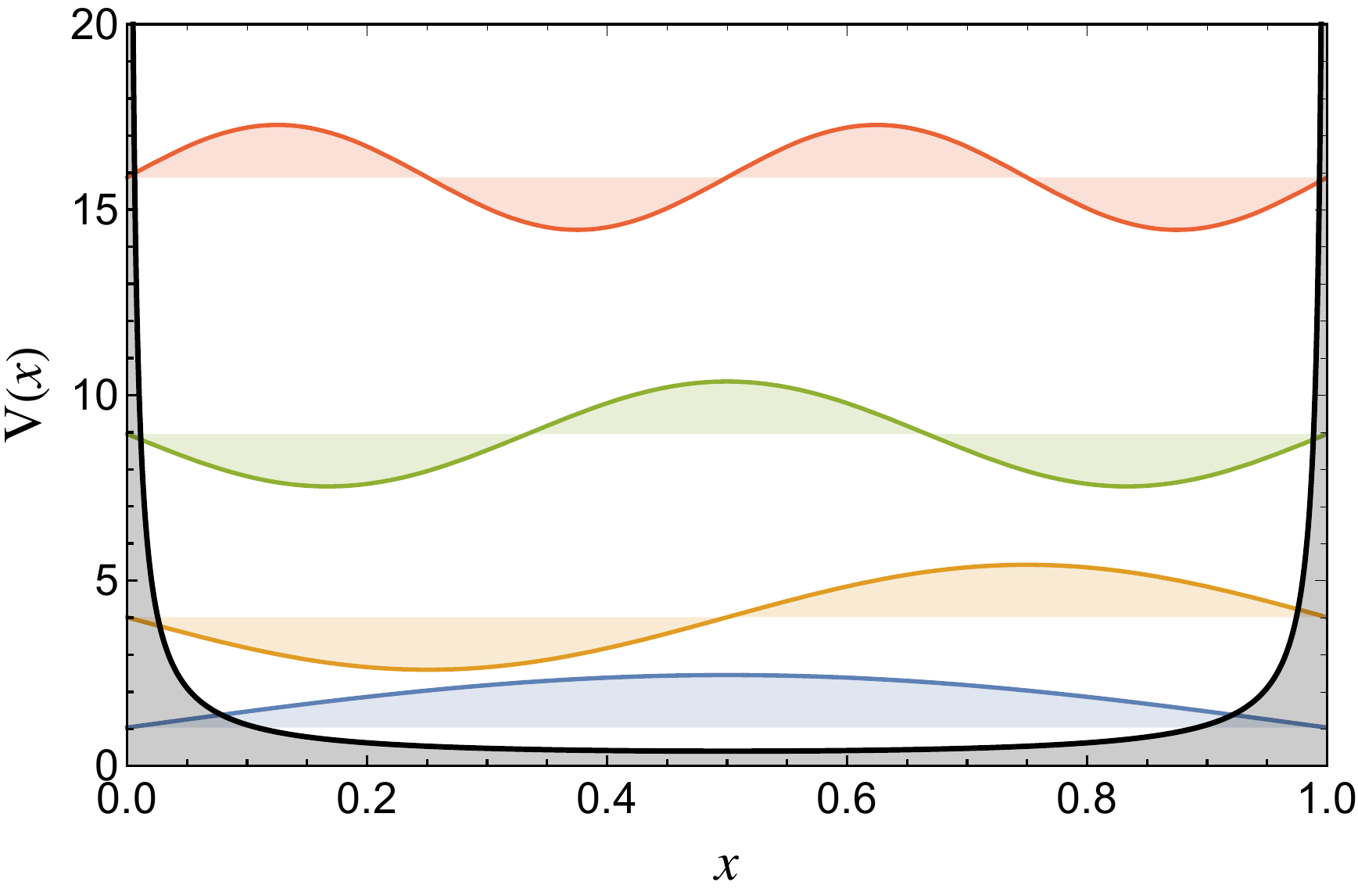}
\caption{Solutions of the SL problem (\ref{eqn:LFSE1D}).}
\label{fig:LFSE1D}
\end{figure}

For $m_q>0$, no analytic solution is known so far, although in special cases, the solution is related to the (regularized) hypergeometric function ${}_2F_1$ \cite{AS, DLMF}. Numerical solutions are shown in Fig.~\ref{fig:LFSE1D}. The $n$-th eigenvalue, again, scales as $\lambda_n \propto n^2$. At the endpoints $x\to 0$, the wave function $\chi(x) \sim x$. Since there is no singularity in the integral of the kinetic energy, the quark mass contribution can be estimated as, $\Delta M^2 = \big[\gamma_\textsc{e} + \ln 2\pi - \mathrm{Ci}(2\pi)\big]m_q^2 + O(m_q^4)$, where, $\gamma_\textsc{e}$ is Euler's constant, $\mathrm{Ci}(2\pi) \approx -0.0225607$ is the cosine integral. \cite{AS, DLMF}

\subsection{Case II: $p(x) = \sigma^2 x(1-x)$}

This case is the LMZV model discussed above. 

\subsection{Case III: $p(x) = \sigma^2 x^2(1-x)^2$}
In this case, we choose $p(x) = \sigma^2 x^2(1-x)^2$.  Then the longitudinal LFSWE becomes, 
\begin{equation}
\Big[\frac{m_q^2}{x}+\frac{m^2_{\bar q}}{1-x} - \sigma^2\frac{\dd}{\dd x} x^2(1-x)^2 \frac{\dd}{\dd x}  \Big]\chi(x) = M^2_\| \chi(x). 
\end{equation}
The associated SL problem is,
\begin{equation}
-\big( x^2(1-x)^2\chi'\big)' + \Big(\frac{\beta^2}{4x} +  \frac{\alpha^2}{4(1-x)}\Big) \chi = \lambda \chi, \quad (\chi(0) = \chi(1) = 0)
\end{equation} 
The analytical solutions are not known. By matching to the singularities, it can be shown that at the endpoints, the wave functions
scales as $\chi(x) \sim \exp\big[-\frac{2}{\sigma}\big(\frac{m_q}{\sqrt{x}}+\frac{m_{\bar q}}{\sqrt{1-x}}\big)\big]$. Numerical solutions show the eigenvalues $\lambda_n \propto n$ for low $n$ and $\lambda_n \propto n^2$ for large $n$. 

\subsection{Case IV: $p(x) = \sigma^2 x^3(1-x)^3$}
In this case, we choose $p(x) = \sigma^2 x^3(1-x)^3$.  Then the longitudinal LFSWE becomes, 
\begin{equation}
\Big[\frac{m_q^2}{x}+\frac{m^2_{\bar q}}{1-x} - \sigma^2\frac{\dd}{\dd x} x^3(1-x)^3 \frac{\dd}{\dd x}  \Big]\chi(x) = M^2_\| \chi(x). 
\end{equation}
The associated SL problem is,
\begin{equation}
-\big( x^3(1-x)^3\chi'\big)' + \Big(\frac{\beta^2}{4x} +  \frac{\alpha^2}{4(1-x)}\Big) \chi = \lambda \chi, \quad (\chi(0) = \chi(1) = 0)
\end{equation} 
The analytical solutions are not known. By matching to the singularities, it can be shown that at the endpoints, the wave functions
scale as $\chi(x) \sim \exp\big[-\frac{1}{\sigma}\big(\frac{m_q}{x}+\frac{m_{\bar q}}{1-x}\big)\big]$. Numerical calculations show that the eigenvalues $\lambda_n \propto n$. 

This case is closely related to the G\l{}azek-Trawi\'nski potential introduced above. To see this, we can express the longitudinal coordinate $\zeta_\| = i\nabla_{\kappa_\|}$ of G\l{}azek and Trawi\'nski 
in terms of $\tilde z = i\partial_x$. Recall, in the equal mass case, $\kappa_\| = m_q \frac{2x-1}{\sqrt{x(1-x)}}$, or equivalently, $x = (1/2)(1+\kappa_\|/\sqrt{4m_q^2+\kappa^2_\|})$. Then, we can obtain $\zeta_\| = (2/m_q)\sqrt{x^3(1-x)^3}i\partial_x$. Therefore, the longitudinal confinement in this model becomes, 
\begin{equation}
V_\|^{(\text{GT})} = -\frac{4\lambda^2}{m_q^2} \partial_x x^3(1-x)^3 \partial_x.
\end{equation}
The confining strength $\sigma = \frac{2\lambda}{m_q}$. With this identification, the wave function at the endpoints becomes, 
$\chi(x) \sim \exp\big(-\frac{m_q^2}{2\lambda x(1-x)}\big)$, in agreement with G\l{}azek and Trawi\'nski's wave function.

There is a caveat here since $\kappa_\perp = \vec k_\perp/\sqrt{x(1-x)}$ contains the longitudinal coordinate $x$. Therefore, 
$\vec k_\perp = \vec\kappa_\perp/\sqrt{4+\kappa^2_\|/m_q^2}$ also depending on $\kappa_\|$. To obtain the 3rd coordinate $\zeta_\|$, we can apply the 
chain rule:
\begin{align}
\zeta_\| \equiv\,& i \frac{\partial}{\partial \kappa_\|} \\
=\,& \frac{\partial\vec k_\perp}{\partial\kappa_\|} \cdot \vec r_\perp + \frac{\partial x}{\partial \kappa_\|} P^+r_+ \\
=\,& - \half(x-\half) x^{\half}(1-x)^{\half} m_q^{-1} \vec k_\perp \cdot \vec r_\perp + 2x^{\half[3]}(1-x)^{\half[3]} m_q^{-1}  \tilde z.
\end{align}

Here $\vec r_\perp = i\partial/\partial{\vec k_\perp}$, $r_+ = i\partial/\partial k^+$, and $P^+r_+ \equiv \tilde z = i\partial_x$ is Brodsky and Miller's frame-independent longitudinal coordinate. Since $\partial \vec k_\perp/\partial \kappa_\| \ne 0$, in general $\zeta_\|$ depends on $\vec r_\perp$. In other words, 
$\vec \kappa$ and $\vec \zeta$ are not independent conjugate pairs, and the longitudinal interaction constructed here may not be exactly the same as the G\l{}azek-Trawi\'nski model.

\begin{table}
\caption{Summary of several longitudinal interactions discussed in this work.}
\label{tab:summary}
\begin{tabular}{c | c | c | c | c c}
\toprule
 interaction & $\tilde z$-space& $M_\text{gs}^2$ & $M^2_n$ & $\chi(x)$ & \\
 \colrule
 $-\sigma^2\partial_x^2$ & $\sigma^2\tilde z^2$ & & $\sim n^2$ & $\sim x(1-x)$ \\
 $-\sigma^2\partial_x x(1-x) \partial_x$ & & $(m_q+m_{\bar q})^2 + \sigma(m_q+m_{\bar q})$ & $\sim n^2$ & $\sim x^\frac{m_{q}}{\sigma}(1-x)^\frac{m_{\bar q}}{\sigma}$ \\
 $-\sigma^2\partial_x x^2(1-x)^2 \partial_x$ & & & $\sim n^2$ & $\sim \exp\big(-\frac{2}{\sigma}\sqrt{\frac{m_q^2}{x}+\frac{m_{\bar q}^2}{1-x}}\big)$ \\
 $-\sigma^2\partial_x x^3(1-x)^3 \partial_x$ & & & $\sim n$ & $\sim \exp\big(-\frac{1}{\sigma}\big({\frac{m_q}{x}+\frac{m_{\bar q}}{1-x}}\big)\big)$ \\
$-g^2\frac{1}{(x-x')^2}$ & $g^2|\tilde z|$ & $\sqrt{\frac{\pi}{3}} g(m_q+m_{\bar q})$ & $\sim n$ & $\sim x^\beta(1-x)^{\beta'}$ \\
\botrule
\end{tabular}
\end{table}

\section{Summary and outlooks}\label{sect:conclusion}

In this work, we discussed the role of longitudinal dynamics in semi-classical light-front Schrödinger wave equations. 
Based on a separation of variables ansatz, we relate the longitudinal confining interaction to the endpoint asymptotics of the resulting wave function as well as to the scaling of the eigenvalues. This enables us to compare longitudinal interactions proposed in the recent literature as well as a broader class of longitudinal interactions based on the Sturm-Liouville theory. We analyzed several concrete cases $V_\| \sim -\partial_x x^\gamma(1-x)^\gamma \partial_x$ ($\gamma=0,1,2,3$). We also pointed out the challenges for constructing valid longitudinal interactions due to the presence of the endpoint singularities in the light front kinetic energy. 

Our motivation is to search for a plausible first approximation to mesons in conjunction with light-front holography. Both the 't~Hooft interaction $\propto\big|\tilde z\big|$ and the LMZV interaction $\propto \partial_x\big(x(1-x)\big)\partial_x$ implement the chiral symmetry breaking for the pion. They both produce power-law-like wave functions at the endpoints. On the other hand, interactions of quadratic form naturally match the transverse holographic confining potential in the nonrelativistic limit. 
We further expect that the resulting wave functions can be used as a basis for the full quantum many-body calculations. In this regard, the LMZV interaction is useful for its computational convenience. Indeed, the longitudinal wave functions from this interaction, the power-law weighted Jacobi polynomials, form a standard basis for wave functions with power-law behavior, such as the solution of the 't~Hooft model. 
Other longitudinal interactions may be useful in different contexts. 

We focused on the light mesons. The reason is that the hyperfine structure of the heavy systems, including the heavy-light mesons are dominated by the one-gluon-exchange interaction \cite{Li:2015zda, Godfrey:1985xj}. A single confining interaction is not sufficient. Indeed, the works on heavy systems have to introduce additional parameters. Moreover, the restoration of the 3D rotational symmetry in the non-relativistic limit puts strong constraints on the confining interactions.\footnote{Here we only retain those longitudinal confining forms that reduce to the harmonic oscillator in the non-relativistic limit due to another symmetry constraint: we require factorization of the CM motion from the internal motion.}
As such, different forms of the longitudinal confinement become nearly identical. 

One of the interesting questions is how to generalize the present investigation to the multi-parton sectors, including notably the baryons. There have been several proposals in the literature, including the pairwise longitudinal interactions \cite{Mondal:2019jdg, Xu:2021wwj}, the quark-diquark model \cite{Yu:2019, Ahmady:2021yzh}, and the orthogonal basis over a $k$-simplex (triangle, tetrahedron, ...) \cite{Chabysheva:2013oka, Chabysheva:2014rra, Burkardt:2016ffk, Shuryak:2022thi}. 

\section*{Acknowledgements}

We wish to thank S.J. Brodsky and G.F. de T\'eramond for enlightening discussions and for their critical reading of an earlier version of the manuscript. We also gratefully acknowledge fruitful discussions with S.J. Brodsky, G.F. de T\'eramond, G.A. Miller, X. Zhao, C. Mondal, J.~Hiller, R.~Sandapen and Q. Wang. This work is supported in part by the Department of Energy under Grants No. DE-FG02-87ER40371, and No. DE-SC0018223 (SciDAC4/NUCLEI). Y.L. is supported by the New faculty start-up fund of the University of Science and Technology of China.

\end{document}